\documentclass[11pt]{article}
\usepackage{graphics}
\begin{document}
\vspace*{1mm}
\begin{center}
{\Huge\sf Quantum Energy Teleportation \\[2mm]
between Spin Particles in a Gibbs State}
\end{center}
\vspace{.1in}
\begin{center}
\hspace{.01\linewidth}
\begin{minipage}[h]{0.31\linewidth}

{\Large Michael R. Frey}  \\
Dept.\ of Mathematics \\[-.5mm]
Bucknell University \\[-.5mm]
Lewisburg, PA 17837, USA  \\[-.5mm]
mfrey@bucknell.edu

\end{minipage}
\begin{minipage}[h]{0.31\linewidth}

{\Large Karl Gerlach}  \\
ITT Corporation \\[-.5mm]
2560 Huntington Avenue \\[-.5mm]
Alexandria, VA 22303, USA   \\[-.5mm]
karlgerlach@comcast.net

\end{minipage}
\begin{minipage}[h]{0.31\linewidth}

{\Large Masahiro Hotta}  \\
Graduate School of Science \\[-.5mm]
Tohoku University \\[-.5mm]
Sendai 980-8578, Japan   \\[-.5mm]
hotta@tuhep.phys.tohoku.ac.jp

\end{minipage}
\end{center}

\vspace*{10mm}
\begin{center}
{\small
{\large\bf ABSTRACT}
\vspace*{-3mm}
 
\begin{quote}
Energy in a multipartite quantum system appears from an operational perspective
to be distributed to some extent non-locally because of correlations extant
among the system's components. This non-locality allows users to transfer,
in effect, locally accessible energy between sites of different system components by LOCC (local
operations and classical communication). Quantum energy teleportation is a three-step LOCC
protocol, accomplished without an external energy carrier, for effectively transferring
energy between two physically separated, but correlated, sites. We apply this LOCC
teleportation protocol to a model Heisenberg spin particle pair initially in a quantum
thermal Gibbs state, making temperature an explicit parameter. We find in this setting that energy
teleportation is possible at any temperature, even at temperatures above the threshold where
the particles' entanglement vanishes. This shows for Gibbs spin states that entanglement is
not fundamentally necessary for energy teleportation; correlation other than entanglement
can suffice. Dissonance---quantum correlation in separable states---is in this regard shown
to be a quantum resource for energy teleportation, more dissonance being consistently
associated with greater energy yield. We compare energy teleportation from particle A to B
in Gibbs states with direct local energy extraction by a general quantum operation on B and
find a temperature threshold below which energy extraction by a local operation is impossible.
This threshold delineates essentially two regimes: a high temperature regime where entanglement
vanishes and the teleportation generated by other quantum correlations yields only vanishingly
little energy relative to local extraction and a second low-temperature teleportation regime
where energy is available at B \emph{only} by teleportation.
\vspace*{4mm}

\end{quote}}
\end{center}
\vspace*{2mm}

\newpage
\begin{center}
{\large\bf I. INTRODUCTION}
\end{center}
\vspace*{-5mm}

Standard quantum teleportation is the transfer of a system's unknown quantum state to a
``blank'' second system \cite{benn}. This is tantamount to a transfer of the system itself
since a system is identified by its quantum state. Quantum teleportation is accomplished
using only local operations and classical communication (LOCC), but it requires quantum
correlation---entanglement---between the two systems. In contrast to state teleportation,
quantum energy teleportation (QET) is effectively a transfer by LOCC of energy between
two components of a multipartite system \cite{hot1,hot2,hot3}. More specifically, QET is
a transfer of local energy, where local (or locally available) energy is energy that can
be extracted from a system component by a local operation. Like state teleportation, QET
relies for its operation on some correlation between the two components, and to users sited
at the two components, energy appears from an operational perspective at least partially
non-locally identified because of
these correlations. For example, zero-point energy, while existing everywhere in a multi-body
system in the ground state, cannot be extracted by local operation at any single site, and
any attempt to do so would only inject additional energy into the system. Thus zero-point
energy is not locally available. On the other hand, a LOCC protocol executed by two users
at the sites of different subsystems can draw zero-point energy from one subsystem by
injecting additional energy into the second subsystem. This is the basis of QET.

A QET protocol was theoretically demonstrated first for spin chains and quantum fields
\cite{hot1,hot2} and subsequently for an elementary ``minimal'' physical model \cite{hot7}.
This minimal model involves just a single maximally entangled spin-$\frac{1}{2}$ particle
pair and a nondemolition measurement of the interaction Hamiltonian. These various
demonstrations show that, by injecting energy at site A, energy can be extracted at site B,
with no external energy carrier. The speed of extraction is limited by that of classical
communication between the sites, consistent with causality. And, because QET only increases
the ratio of locally available energy to the total energy in the energy extraction region,
with no change to the total amount of energy, QET strictly conserves local energy
\cite{hot1,hot2,hot3}. A proposal has been made to use edge channel currents in a quantum
Hall system to experimentally verify QET \cite{hotexp}.

Energy teleportation has a range of implications for fundamental physics. For
example, it suggests local energy density fluctuation as a way to address
entanglement in condensed matter systems. QET may in this regard constitute
a new tool for a quantum Maxwell's demon, allowing the demon to observe and
react to local quantum fluctuations of an interacting many-body system at zero
temperature. Past works on quantum demons assume that interactions among
observed subsystems are negligbly small, eliminating a direct role for ground
state entanglement \cite{d1,d2,d3}. A demon equipped to perform QET can by
indirect measurement exploit ground-state entanglement to extract work,
potentially opening the way to a new paradigm for quantum information
thermodynamics. Also, QET bears on local cooling in quantum many-body systems.
Local measurement of zero-point fluctuation on a subsystem generally injects
some energy, resulting in an excited state. We then naturally ask whether all
the injected energy can be retrieved using only local operations on the
measured subsystem. With the perspective of QET, the answer is no; some
residual energy unavoidably remains in the system from any local-cooling
procedure \cite{hot1}. This is because the local measurement breaks a part of
the ground-state entanglement and the broken entanglement cannot be restored
by local operations. In fact, the residual energy is lower bounded by the
total amount of energy that can be teleported by use of the information from
the local measurement \cite{hot4}. QET considered in the setting of black hole
physics provides a new method \cite{hotblack} analogous to Hawking radiation
\cite{hawk} for reducing the area of the event horizon. Consider a quantum
field measurement outside a massive black hole that provides information about
quantum fluctuations. Positive-energy wave packets of the field are generated
during the measurement (based on approximating the quantum field's
pre-measurement state by a Minkowski vacuum state and making a passivity
argument). Suppose that the black hole absorbs these wave packets. Then,
significantly, part of the absorbed energy outside the horizon can be
retrieved by QET. Using the measurement information,
negative energy wave packets can be generated outside the horizon by
extracting positive energy out of the zero-point fluctuation of the fields.
The negative energy of the wave packets propagates across the event horizon
and may pair-annihilate with positive energy of matter falling inside the
black hole. This process is akin to spontaneous emission of Hawking radiation
or, as it often called, black hole tunneling \cite{hawk,pari}. The net effect
of this process is to decrease the horizon area, which is proportional to the
black hole entropy. This result may from an information theory viewpoint
clarify the origin of black hole entropy. QET appears by these examples to be
a fundamental physical process relevant to different branches of physics.

We study QET in this paper within the framework of a coupled pair of spin-$\frac{1}{2}$
particles, focusing on the quantum Gibbs states of the particle pair. To locally interact
with a single particle in the pair for the purpose of QET, we adopt the same nondemolition
measurement of the interaction Hamiltonian used in \cite{hot7}. To physically motivate
the Gibbs states, we assume that the particles are coupled to bosonic environments
(thermal baths) with interactions that are very weak relative to the interaction between
the two particles. This assumption (with others) allows Gibbs states to be interpreted
as a physical thermal state with equilibrium temperature as a parameter \cite{thermo}.
Our study of these states introduces their (equilibrium) temperature as an explicit
parameter, allowing us to investigate for this model 1) the extent to which temperature
restricts QET, 2) the role of different forms of quantum correlation in QET and 3) the
performance of QET relative to direct local energy extraction. In this
investigation we show that energy teleportation is possible for any Gibbs state of the
particle pair, establishing in principle that QET can be accomplished under suitable
conditions at any temperature. We observe that in our Gibbs states the particles' spins
are quantum correlated (in the sense of quantum discord) at all temperatures, though
significantly they are entangled only at temperatures below a certain threshold. We
conclude from this that in Gibbs states the correlation essential for energy
teleportation need not fundamentally be entanglement---correlation beyond entanglement
can support energy teleportation. This adds to a growing number of applications in which
quantum dissonance (discord without entanglement) is demonstrably a quantum resource
\cite{roa,datt,mod2,frey,dat2}. Finally in this study, to better understand QET and
underscore its unique capability, we compare QET from particle A to B with direct local
energy extraction by a general quantum operation on B. Concerning local extraction of
energy, we obtain two interesting results: 1) no energy can be extracted at B by a local
unitary operation at any temperature but 2) local energy extraction at B is possible by
a general (Kraus operator-sum) quantum operation, provided the temperature is above a
threshold. This parametric threshold marks two temperature regimes: a high temperature
regime where teleportation yields only vanishingly small amounts of energy relative to
local extraction and a low-temperature teleportation regime where energy is available at B
\emph{only} by teleportation. These regimes indicate by their nature that some quantum
correlation---entanglement or otherwise---is effectively required for Gibbs state QET.

The remainder of the paper is organized as follows. Section II introduces our quantum
model of two spin-$\frac{1}{2}$ particles. In this section we focus on the Gibbs states
associated with this model, identifying the type and degree of the quantum correlation
within these states as a function of temperature. In section III we specify the QET
protocol for teleporting energy from particle A to B and establish the central result
that the protocol yields a positive amount of energy at the site of B, doing so at any
temperature. In section IV we study energy extraction at B by local operations and
contrast this with QET. We conclude in section V with some last remarks, including a
discussion of dissonance as a quantum resource for Gibbs state QET.

\vspace*{+2mm}
\begin{center}
{\large\bf II. TWO-PARTICLE SYSTEM}
\end{center}
\vspace*{-3mm}

We consider a model Heisenberg spin-$\frac{1}{2}$ particle pair, focusing on the Gibbs
states associated with this model. These Gibbs states include as a special case the
maximally entangled ground state in the QET study in \cite{hot4}, they have a ready physical
motivation and they are a frequent vehicle for studies of entanglement and discord in
spin systems \cite{wan2,asou,zhan,wer1,xin,pal}. In this section we quantify the type
and degree of the quantum correlation within these states. In particular, we derive an
expression for the Gibbs states' quantum discord, such discord being called conventionally
thermal discord. Quantum discord quantifies the presence of quantum correlation broadly
defined, and we find positive thermal discord at all finite temperatures, across the whole
class of Gibbs states, even in those Gibbs states without entanglement.

\underline{Model}: Consider two spin-$\frac{1}{2}$ particles, A and B, with Hamiltonian
\begin{equation}
\mbox{\bf H} = \mbox{\bf H}_A \otimes \mbox{\bf I} +  \mbox{\bf I}
\otimes \mbox{\bf H}_B + \mbox{\bf V}
\label{eq:ham}
\end{equation}
\noindent
where
\begin{eqnarray}
&& \mbox{\bf H}_A = \mbox{\bf H}_B = \frac{1}{m} \mbox{\bf I} + \mbox{\boldmath$\sigma$}_z \, ,  \nonumber \\
\label{eq:comp} \\[-5mm]
&& \!\!\!\!\!\!\!\!\!
\mbox{\bf V} = 2\kappa \; \mbox{\boldmath$\sigma$}_x \otimes \mbox{\boldmath$\sigma$}_x
+ 2\frac{\kappa^2}{m} \mbox{\bf I} \otimes \mbox{\bf I}
\nonumber
\end{eqnarray}
\noindent
with $m=\sqrt{1+\kappa^2}$ and Pauli operators $\mbox{\boldmath$\sigma$}_x$,
$\mbox{\boldmath$\sigma$}_y$, $\mbox{\boldmath$\sigma$}_z$.
The particle pair model (\ref{eq:ham}) and (\ref{eq:comp}) is Hotta's minimal
model \cite{hot4} with one independent parameter $\kappa \geq 0$ and dimensionless
energy. In the components (\ref{eq:comp}) of $\mbox{\bf H}$, the constants---those
terms involving identity operators $\mbox{\bf I}$---do not change the relative
magnitudes of the eigenenergies of $\mbox{\bf H}$; they just serve to set
the ground eigenenergy to $E_0=0$ and we include them here for consistency
with \cite{hot4}. The Hamiltonian (\ref{eq:ham}) is equivalently that of a
two-qubit Ising spin chain in a transverse magnetic field; viewed so, $\kappa$
is the strength of the spin coupling relative to that of the magnetic field.
The two-qubit system with Hamiltonian (\ref{eq:ham}) has eigenenergies
\begin{equation}
E_0 = 0 \; , \;\;\;   E_1 = 2m - 2\kappa \; , \;\;\;
E_2 = 2m + 2\kappa \; , \;\;\;   E_3 = 4m
\label{eq:en}
\end{equation}
\noindent
and corresponding eigenstates
\begin{eqnarray}
&& \!\!\!\!\!\!\!\!\!\!\!\!
|E_0\rangle= \sqrt{\frac{m-1}{2m}}\,|00\rangle
-\sqrt{\frac{m+1}{2m}}\,|11\rangle , \,
\nonumber \\
&& \;\;\;\;
|E_1\rangle= \frac{|01\rangle-|10\rangle}{\sqrt{2}} , \,
\nonumber \\
&& \;\;\;\;
|E_2\rangle= \frac{|01\rangle+|10\rangle}{\sqrt{2}} , \,
\nonumber \\
&& \!\!\!\!\!\!\!\!\!\!\!\!
|E_3\rangle= \sqrt{\frac{m+1}{2m}}\,|00\rangle
+\sqrt{\frac{m-1}{2m}}\,|11\rangle , \,
\nonumber
\end{eqnarray}
\noindent
in the uncoupled basis $\{ |00\rangle, |01\rangle, |10\rangle, |11\rangle\}$.
Note that $E_0 < E_1 \leq E_2 < E_3$ since $m>\kappa$.

\underline{Gibbs states}: Suppose each particle in the model pair is weakly
coupled (Born approximation) with its own bosonic heat bath at temperature $T$.
Then the eigenstates' canonical occupation (Gibbs) probabilities for the
particle pair are, for $i=0,1,2,3$,
\begin{equation}
p_i(T) = \frac{1}{Z} \exp\left( - \frac{E_i}{kT}\right)
\label{eq:gibb}
\end{equation}
\noindent
where $k$ is Boltzmann's constant, and $Z$ is the partition function
\begin{eqnarray}
&& \!\!\!\!\!\! Z = \sum_{j=0}^3 \exp\left( -\frac{E_j}{kT}
\right)     \nonumber \\
&&  = 2\exp\left( -\frac{2m}{kT} \right)
\left( \cosh\frac{2m}{kT} +  \cosh\frac{2\kappa}{kT} \right) . \nonumber
\end{eqnarray}
\noindent
The quantum state $\rho(T)$ of the particle pair in thermal equilibrium
at temperature $T$ is therefore
\begin{equation}
\rho(T) = \sum_{i=0}^3 p_i(T) |E_i\rangle\langle E_i|
= \frac{1}{4} \left( \matrix{ 1+c_3-2r & 0 & 0 & -c_1-c_2 \cr
0 & 1-c_3 & -c_1+c_2 & 0 \cr
0 & -c_1+c_2 & 1-c_3 & 0 \cr
-c_1-c_2 & 0 & 0 & 1+c_3+2r}
\right)  \label{eq:ther}
\end{equation}
\noindent
where
\begin{equation}
c_1 = \frac{2}{mZ}\exp\left(-\frac{2m}{kT}\right)\left(
m\sinh\frac{2\kappa}{kT} + \kappa\sinh\frac{2m}{kT} \right)    ,
\label{eq:c1}
\end{equation}
\begin{displaymath}
c_2 = \frac{2}{mZ}\exp\left(-\frac{2m}{kT}\right)\left(
-m\sinh\frac{2\kappa}{kT} + \kappa\sinh\frac{2m}{kT} \right)   ,
\end{displaymath}
\begin{equation}
c_3 = \frac{4}{Z}\exp\left(-\frac{2m}{kT}\right)
\sinh\frac{m+\kappa}{kT}\sinh\frac{m-\kappa}{kT}
\label{eq:c3}   ,
\end{equation}
\begin{equation}
r = \frac{2}{mZ}\exp\left(-\frac{2m}{kT}\right)
\sinh\frac{2m}{kT}     .
\label{eq:rr}
\end{equation}
\noindent
In particular, we recover from (\ref{eq:ther}) that $\rho(0)$ and
$\rho(\infty)$ are, respectively, the ground state $|E_0\rangle\langle E_0|$
and the completely mixed state $\frac{1}{4} \mbox{\bf I}\otimes\mbox{\bf I}$.

\underline{Thermal discord}: The total correlation, both quantum and classical,
in a bipartite system in a quantum state $\omega$ is quantified by the quantum
mutual information \cite{slou}, which is given by
\begin{equation}
I[\omega]=S(\omega_A)+S(\omega_B)-S(\omega)
\label{eq:info}
\end{equation}
\noindent
where $\omega_A=\mbox{tr}_B[\omega]$ and $\omega_B=\mbox{tr}_A[\omega]$ are the
marginal states of parts A and B of the system and $S(\cdot)$ is von Neumann
entropy \cite{niel}. For a qubit pair in the Gibbs state (\ref{eq:ther}),
the joint and marginal entropies (in bits) of the pair are
\begin{equation}
S(\rho(T)) = -\sum_{i=0}^3 p_i(T)\log_2 p_i(T)
\label{eq:joint}
\end{equation}
\noindent
and
\begin{equation}
S(\rho_A(T)) = S(\rho_B(T)) = h(r)
\label{eq:marg}
\end{equation}
\noindent
where in (\ref{eq:joint}) the $p_i(T)$ are the Gibbs probabilities
(\ref{eq:gibb}), and $r$ in (\ref{eq:marg}) is given by (\ref{eq:rr}) with
\begin{equation}
h(x) = \frac{1+x}{2}\log_2\frac{2}{1+x}+\frac{1-x}{2}\log_2\frac{2}{1-x} \, .
\label{eq:hx}
\end{equation}
\noindent
The quantum mutual information in a qubit pair in state $\rho(T)$ is, from
(\ref{eq:joint}) and (\ref{eq:marg}) and after some calculation,
\begin{equation}
I[\rho(T)] = 2h(r) - \log_2 Z - \frac{\langle \mbox{\bf H} \rangle}{kT} \log_2 e
\label{eq:tot}
\end{equation}
\noindent
where $\langle \mbox{\bf H} \rangle = \mbox{tr}[\mbox{\bf H}\rho(T)]$ is the
average energy of the particle pair in the thermal state $\rho(T)$.

The classical part of the total correlation (\ref{eq:info}) in parts A and B
of a bipartite quantum system is defined to be the reduced uncertainty about
the state of, say, A by measurement of B \cite{olli,hend}. Suppose we make a
von Neumann measurement $\{ \mbox{\bf M}_k\}$ of B with one-dimensional projectors
$\mbox{\bf M}_k$ such that $\sum_k \mbox{\bf M}_k = \mbox{\bf I}$. This measurement
casts the bipartite system, originally in state $\omega$, into the state
\begin{displaymath}
\omega_k = \frac{1}{q_k} (\mbox{\bf I} \otimes \mbox{\bf M}_k)\,\omega
\,(\mbox{\bf I} \otimes \mbox{\bf M}_k)
\end{displaymath}
\noindent
with probability $q_k = \mbox{tr}[ (\mbox{\bf I} \otimes \mbox{\bf M}_k)\omega
(\mbox{\bf I} \otimes \mbox{\bf M}_k) ]$.
Depending on the measurement outcome, the reduction in uncertainty about the
state of A is $S(\omega_A)-S(\omega_k)$, with average reduction
$S(\omega_A)-\sum_k q_k S(\omega_k)$. The supremum of this average reduction
through measuring B is defined to be the classical part
\begin{equation}
C[\omega] = \sup\limits_{\{\mbox{\scriptsize\bf M}_k\}}
\left( S(\omega_A)-\sum_k q_k S(\omega_k) \right)
\label{eq:clas}
\end{equation}
\noindent
of the total correlation in $\omega$. The optimization in (\ref{eq:clas}) is more
generally taken over quantum measurements described by positive operator-valued
measures, but for two-qubit states the optimal measurement is known to be
projective \cite{hami}. Definition (\ref{eq:clas}), involving as it does
measurement of subsystem B of the bipartite system, is not symmetrical in
A and B and, in fact, the two possible versions of $C[\omega]$ are generally
not equal \cite{olli,feld}. This is not a present concern, though, because $\rho(T)$
is qubit exchange symmetric.

The supremum in the definition (\ref{eq:clas}) of classical correlation is readily found
analytically for two-qubit Bell-diagonal states \cite{lang}. Beyond the Bell-diagonals,
though, the supremum presents a much greater challenge, and there are some erred results
in the literature \cite{huang}. The Gibbs states $\rho(T)$ in (\ref{eq:ther}) are not
Bell-diagonal. Instead, they are a subset of the broader class of states studied in
\cite{wer1,werl,qion}, where in each of these studies the classical correlation was
found by numerical search. Here, expressions (\ref{eq:c1}), (\ref{eq:c3}), and
(\ref{eq:rr}) for $c_1$, $c_3$, and $r$ constrain the form of $\rho(T)$ in (\ref{eq:ther})
sufficiently to allow us to determine the supremum in (\ref{eq:clas}) and obtain an
analytical expression for the classical correlation. We find
\begin{equation}
C[\rho(T)] = h(r)-h\Big(\sqrt{r^2+c_1^2}\,\Big)  \, .
\label{eq:our}
\end{equation}
\noindent
Details of the calculation of (\ref{eq:our}) are given in Appendix A.

The difference between the total correlation in $\rho(T)$ in (\ref{eq:tot}) and
its classical correlation in (\ref{eq:our}) is the quantum discord $D[\rho(T)]$.
Quantum discord quantifies the quantum correlation, entanglement and otherwise,
in a bipartite state \cite{olli,hend}, and it has different important operational
interpretations \cite{cava,madh,stre,pian} to support its use for this purpose.
When the state is separable, any nonzero discord is due strictly to quantum
correlation other than entanglement. Discord in Gibbs states is commonly called
thermal discord. Also, following Modi et al.\ \cite{modi}, we call positive
discord in the absence of entanglement dissonance and say that a separable state
with positive discord is dissonant. From (\ref{eq:info}) and (\ref{eq:our}) the
discord in $\rho(T)$ is
\begin{equation}
D[\rho(T)] = h(r)+h\Big(\sqrt{r^2+c_1^2}\,\Big)
\log_2 Z - \frac{\langle \mbox{\bf H} \rangle}{kT} \log_2 e \, .
\label{eq:disc}
\end{equation}
\noindent
The ground thermal state $\rho(0)=|E_0\rangle\langle E_0|$ is pure so its quantum
correlation is solely entanglement, and its discord (\ref{eq:disc}) is the entropy
of entanglement $D[\rho(0)]=S(\rho_A(T))=h(r)$. Figure 1 shows for chosen values of
$\kappa$ that $D[\rho(T)]$ is positive and decreasing with temperature $T$ for all $T$.

\begin{figure}[t]
\centerline{\scalebox{.9}{\includegraphics{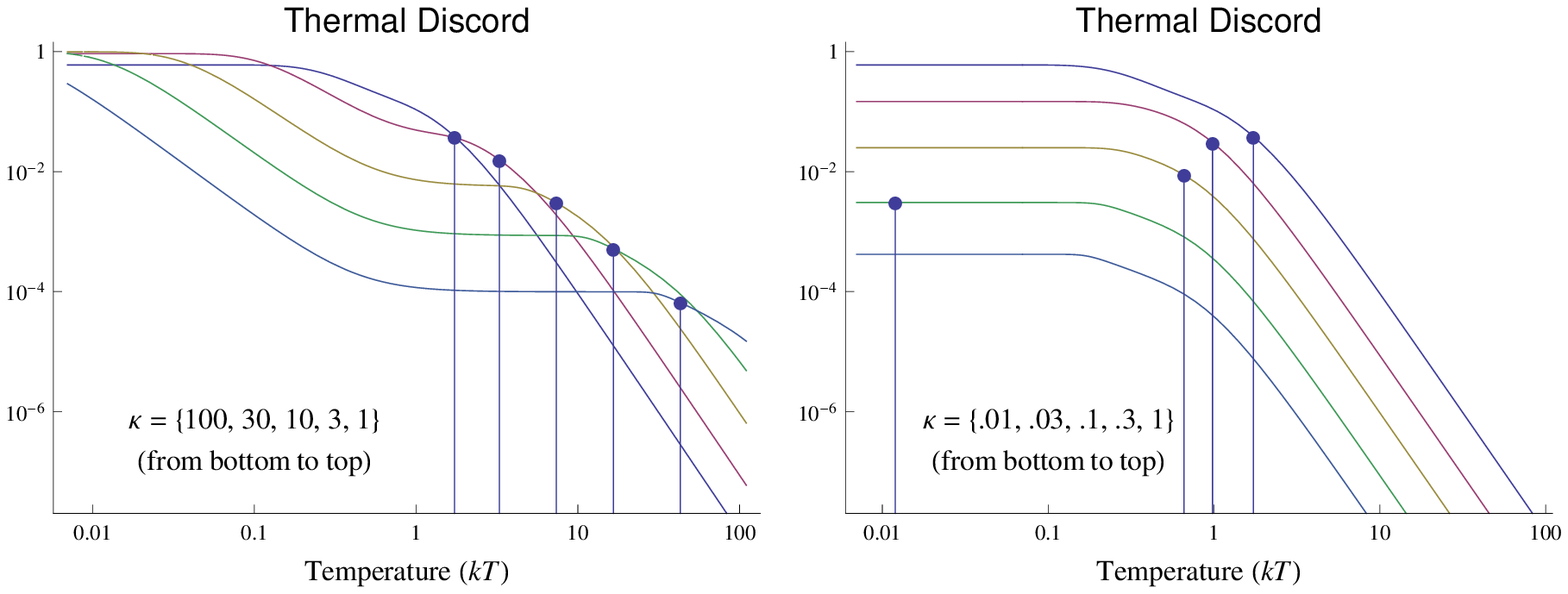}}}
\vspace*{8mm}
\begin{center}
\parbox{4.9in}{\small
Figure 1. Thermal discord in the state $\rho(T)$. The left panel shows cases of
the particle pair model with $\kappa \geq 1$, and the right panel shows cases
with $\kappa \leq 1$.  The state $\rho(T)$ is separable for temperatures
above $T_e$, to the right of $\bullet$ on each curve.}
\end{center}
\vspace*{5mm}
\end{figure}
 
\underline{Entanglement}: Entanglement is a particular form of quantum correlation,
readily detected in the Gibbs states $\rho(T)$ by the PPT criterion \cite{pere}.
The PPT criterion yields (see Appendix B) that $\rho(T)$ is separable if and only if
\begin{equation}
m\cosh\frac{2\kappa}{kT} \geq \kappa\sinh\frac{2m}{kT}   \, .
\label{eq:cond}
\end{equation}
\noindent
This condition is saturated by a critical temperature $T=T_e$, which increases with the
coupling $\kappa$. Below $T_e$ (in the shaded region in Fig. 5) the particles are
entangled to some degree, while at and above $T_e$ there is zero entanglement and
$\rho(T)$ is separable. A temperature threshold for entanglement is usual for thermal
spin-$\frac{1}{2}$ systems \cite{osen}.

\begin{center}
\vspace*{2mm}
{\large\bf III. ENERGY TELEPORTATION}
\end{center}
\vspace*{-2mm}

We now apply the QET protocol to our model particle pair in the thermal state
$\rho(T)$ given by (\ref{eq:ther}), to teleport energy from the site of particle
A to that of B. The QET protocol is known \cite{hot4} to accomplish this in the
zero-temperature case $\rho(0)=|E_0\rangle\langle E_0|$. We show that the
QET protocol succeeds with {\em any} Gibbs state $\rho(T)$, even those states
without entanglement.

The QET protocol proceeds in three steps. Step I is a measurement of particle
A's observable $\mbox{\boldmath$\sigma$}_x$. This local measurement on A,
independent of B, is a nondemolition measurement in as much as the observable
$\mbox{\boldmath$\sigma$}_x$ commutes with the interaction component
(\ref{eq:comp}) of the system Hamiltonian,
$[ \mbox{\boldmath$\sigma$}_x\otimes \mbox{\bf I},\mbox{\bf V}]=0$.
This measurement has the key effect of moving the particle pair to a new
state with changed average local and total energies. Step II of
the protocol is to classically communicate the outcome $\alpha=\pm1$ of the step I
measurement to the site of particle B. In step III, at the site of B, the communicated
outcome $\alpha$ is used to choose a unitary operation $\mbox{\bf U}(\alpha)$ for
local application. This local unitary operation changes, again, the average energy
of the particle pair. A gain $E_{A}>0$ in system energy in step I indicates that the
measurement device at site A has deposited energy into the particle pair, while a
system energy loss $E_{B}>0$ in step III indicates that energy is extracted at site
B from the particle pair. These energy changes combine to achieve the effect of
energy transport from site A to site B. This effect is termed energy teleportation
because 1) being limited only by the communication speed in step II, it can be
accomplished faster than the energy diffusion velocity within the system, 2) A and B
can be a physical distance apart, and 3) no external energy carrier is involved. To
show that the protocol succeeds for any Gibbs state $\rho(T)$ of our qubit pair, we
check that the particle pair's energy gain $E_{A}$ in step I and energy loss $E_{B}$
in step III are both positive. This QET protocol is superficially similar to remote
state preparation (RSP). In each of QET and RSP the aim is to prepare subsystem B by
operation on subsystem A, and each relies on correlation between the two subsystems.
The particular aim of RSP, though, is to impose a known state on the target subsystem,
measuring success by the fidelity of the schieved state \cite{ben2}. And to accomplish
this, the RSP protocol depends on the specific quantum state to be imposed. By contrast,
rather than aim to establish a particular state, QET seeks to maximize the locally
available energy in subsystem B, and to this end, the QET protocol is designed in accord
with the interaction component of the Hamiltonian, and not directly with any initial
or desired quantum state.

Prior to initiating the QET protocol, the average energy in the particle pair in the
Gibbs state $\rho(T)$ is
\begin{displaymath}
\langle \mbox{\bf H} \rangle = \mbox{tr}[\mbox{\bf H}\rho(T)]
=\sum_{i=o}^3p_i(T)E_i
\end{displaymath}
\noindent
where the $p_i(T)$ are the Gibbs probabilities (\ref{eq:gibb}) and
the $E_i$ are the system energies (\ref{eq:en}). Simple calculation yields
\begin{equation}
\langle \mbox{\bf H} \rangle = 2m - 2\kappa c_1 - 2r
\label{eq:h}
\end{equation}
\noindent
in terms of (\ref{eq:c1}) and (\ref{eq:rr}).
 
Consider measuring particle A as specified by the QET protocol.
The projectors associated with the observable $\mbox{\boldmath$\sigma$}_x$ are
\begin{displaymath}
\Pi(\alpha)=\frac{1}{2}(\mbox{\bf I}+\alpha \mbox{\boldmath$\sigma$}_x)
\end{displaymath}
\noindent
for $\alpha=\pm 1$, and the post-measurement system state is,
depending on $\alpha$,
\begin{equation}
\rho_{\mbox{\scriptsize I}}(T,\alpha)
= \frac{[\Pi(\alpha)\otimes\mbox{\bf I}]\, \rho(T) \,
[\Pi(\alpha)\otimes \mbox{\bf I}]}{q(\alpha)}
\label{eq:stat}
\end{equation}
\noindent
where $q(\alpha)=\mbox{tr}[(\Pi(\alpha)\otimes\mbox{\bf I})\,\rho(T)\, (\Pi(\alpha)
\otimes\mbox{\bf I})]$ is the probability of the outcome $\alpha$.
The energy in the post-measurement state (\ref{eq:stat}) is
\begin{equation}
\langle \mbox{\bf H}_{\mbox{\scriptsize I}}(\alpha) \rangle
=\mbox{tr}[\mbox{\bf H}\rho_{\mbox{\scriptsize I}}(T,\alpha)]
=\frac{1}{q(\alpha)}\sum_{i=0}^3   p_i(T)
\langle E_i|\mbox{\bf H}_{\mbox{\scriptsize I}}(\alpha)|E_i\rangle
\label{eq:ena}
\end{equation}
\noindent
where $\mbox{\bf H}_{\mbox{\scriptsize I}}(\alpha)
=[\Pi(\alpha)\otimes\mbox{\bf I}]\,\mbox{\bf H}\,
[\Pi(\alpha)\otimes\mbox{\bf I}]$. Averaging the energies (\ref{eq:ena}) over
the two measurement outcomes $\alpha=\pm 1$, we have
\begin{displaymath}
\langle \mbox{\bf H}_{\mbox{\scriptsize I}} \rangle = \sum_{i=0}^3 p_i(T) \langle E_i|
(\mbox{\bf H}_{\mbox{\scriptsize I}}(1)+\mbox{\bf H}_{\mbox{\scriptsize I}}(-1)) |E_i\rangle
\, ,
\end{displaymath}
\noindent
which after some calculation is
\begin{equation}
\langle \mbox{\bf H}_{\mbox{\scriptsize I}} \rangle
= 2m - 2\kappa c_1 - r \,  .
\label{eq:hi}
\end{equation}
\noindent
According to (\ref{eq:h}) and (\ref{eq:hi}), the average gain
$E_A = \langle \mbox{\bf H}_{\mbox{\scriptsize I}} \rangle-\langle \mbox{\bf H}\rangle$
in system energy that results from the measurement of particle A is $E_A = r$. The
quantity $r$ can be seen from (\ref{eq:rr}) to be a positive decreasing function of
$T$ for all $\kappa$. Therefore, the QET protocol's measurement of particle A injects
energy into the system on average, injecting more energy for lower temperature.

Now we consider the extraction of energy at the site of particle B. Suppose that
the outcome $\alpha$ of the measurement of particle A has been communicated
to the site of B, and suppose the local unitary operation
\begin{equation}
\mbox{\bf U}(\alpha) = \mbox{\bf I}\cos\theta - i\alpha\mbox{\boldmath$\sigma$}_{\! y}\sin\theta
\label{eq:ua}
\end{equation}
\noindent
specified by step III of the QET protocol is applied to B, where the angle
$\theta$ in (\ref{eq:ua}) is an adjustable real parameter. The state of the
particle pair at the completion of step III is, depending on $\alpha$,
\begin{equation}
\rho_{\mbox{\scriptsize III}}(T,\alpha)
= \frac{[\Pi(\alpha)\otimes\mbox{\bf U}(\alpha)]\, \rho(T) \,
[\Pi(\alpha)\otimes \mbox{\bf U}(\alpha)]}{q(\alpha)}   \, .
\label{eq:stt}
\end{equation}
\noindent
The energy in the state (\ref{eq:stt}) is
\begin{equation}
\langle \mbox{\bf H}_{\mbox{\scriptsize III}}(\alpha) \rangle
=\mbox{tr}[\mbox{\bf H}\rho_{\mbox{\scriptsize III}}(T,\alpha)]
=\frac{1}{q(\alpha)}\sum_{i=0}^3\langle E_i|\mbox{\bf H}_{\mbox{\scriptsize III}}
(\alpha)|E_i\rangle
\label{eq:enb}
\end{equation}
\noindent
where $\mbox{\bf H}_{\mbox{\scriptsize III}}(\alpha)
=[\Pi(\alpha)\otimes\mbox{\bf U}(\alpha)]\,\mbox{\bf H}\,
[\Pi(\alpha)\otimes\mbox{\bf U}(\alpha)]$. Averaging the energies (\ref{eq:enb})
over the two measurement outcomes $\alpha=\pm 1$, we have
\begin{displaymath}
\langle \mbox{\bf H}_{\mbox{\scriptsize III}} \rangle
= \sum_{i=0}^3 p_i(T) \langle E_i|
(\mbox{\bf H}_{\mbox{\scriptsize III}}(1)+\mbox{\bf H}_{\mbox{\scriptsize III}}(-1))
|E_i\rangle     \, .
\end{displaymath}
\noindent
We then calculate that
\begin{equation}
\langle \mbox{\bf H}_{\mbox{\scriptsize III}} \rangle
= 2m+\frac{c_2-c_1}{2}(2\kappa\cos2\theta-\sin2\theta)
-r\left( \kappa\sin2\theta+(m^2+\kappa^2)\cos2\theta \right) \, .
\label{eq:hiii}
\end{equation}
\noindent
Comparing (\ref{eq:hi}) and (\ref{eq:hiii}), we find that the average
loss of energy in the particle pair due to the local unitary operation
$\mbox{\bf U}(\alpha)$ is
\begin{eqnarray}
&& \!\!\!\!\!\!\!\!\!\!\!\!\!\!\!\!\!
E_B(\theta)= \langle \mbox{\bf H}_{\mbox{\scriptsize I}} \rangle
- \langle \mbox{\bf H}_{\mbox{\scriptsize III}} \rangle \nonumber \\
&& = a(\kappa,T)\sin2\theta-b(\kappa,T)(1-\cos2\theta)  \label{eq:extr}
\end{eqnarray}
\noindent
where the coefficients $a(\kappa,T)$, $b(\kappa,T)$ are
\begin{eqnarray}
&& \!\!\!\!\!\!\!\!\!\!\!\!\!\!\!\!\!\!\!\!
a(\kappa,T) = \kappa r +\frac{c_2-c_1}{2}       \nonumber \\
&& =\frac{4\kappa}{ZkT}\exp\left(-\frac{2m}{kT}\right)
\left( s\bigg(\frac{2m}{kT}\bigg)
- s\bigg(\frac{2\kappa}{kT}\bigg) \right) \, , \nonumber \\
&& \!\!\!\!\!\!\!\!\!\!\!\!\!\!\!\!\!\!\!\!
b(\kappa,T) =  (\kappa^2+m^2)r - \kappa (c_2-c_1)    \nonumber \\
&& = \frac{4}{ZkT}\exp\left(-\frac{2m}{kT}\right)
\left( 2\kappa^2 s\bigg(\frac{2\kappa}{kT}\bigg)
+(\kappa^2+m^2)s\bigg(\frac{2m}{kT}\bigg) \right)     \nonumber
\end{eqnarray}
\noindent
with $s(x)=\sinh(x)/x$. The energy $E_B(\theta)$ is the average energy
extracted at site B by the QET protocol, as a function of the
angle $\theta$ used in (\ref{eq:ua}) in step III. The optimal choice
$\theta=\theta_o$ to maximize $E_B(\theta)$ is, from (\ref{eq:extr}), given by
\begin{displaymath}
\tan2\theta_o = \frac{a(\kappa,T)}{b(\kappa,T)} \, .
\end{displaymath}
\noindent
Substituting $\theta_o$ into (\ref{eq:extr}), we find that the maximum
extracted energy $E_B=E_B(\theta_o)$ at site B with the QET protocol is
\begin{equation}
E_B = \sqrt{a(\kappa,T)^2+b(\kappa,T)^2} - b(\kappa,T) .
\label{eq:max}
\end{equation}
\noindent
Plots of the maximum extracted energy $E_B$ for different $\kappa$ in Fig.\ 2 show
that $E_B$ is a decreasing function of temperature, with a temperature threshold for the
decrease for $\kappa <1$ (weak spin coupling). In the regime $\kappa <1$ where this
threshold exists, the temperature of the particle pair can be increased up to the
threshold with almost no decrease in teleported energy. Figure 2 also indicates that
maximum energy teleportation occurs with $\kappa \approx 1$; that is, when the strength
of the particles' coupling is comparable to that of the external magnetic field.

Significantly for what we wish to establish, we see in all cases in Fig.\ 2 that
$E_B$ is positive. In fact, inspection of (\ref{eq:max}) shows immediately that $E_B$
is positive for all finite temperatures $T$ and non-zero particle couplings $\kappa$.
This establishes one of our central results: in the setting of our two-particle model,
the QET protocol yields a positive amount of energy at site B. It does so at any
temperature and across the whole family of spin particle pair systems parameterized by
$\kappa>0$, with maximum teleported energy $E_B$ given by (\ref{eq:max}).

\begin{figure}[t]
\centerline{\scalebox{0.9}{\includegraphics{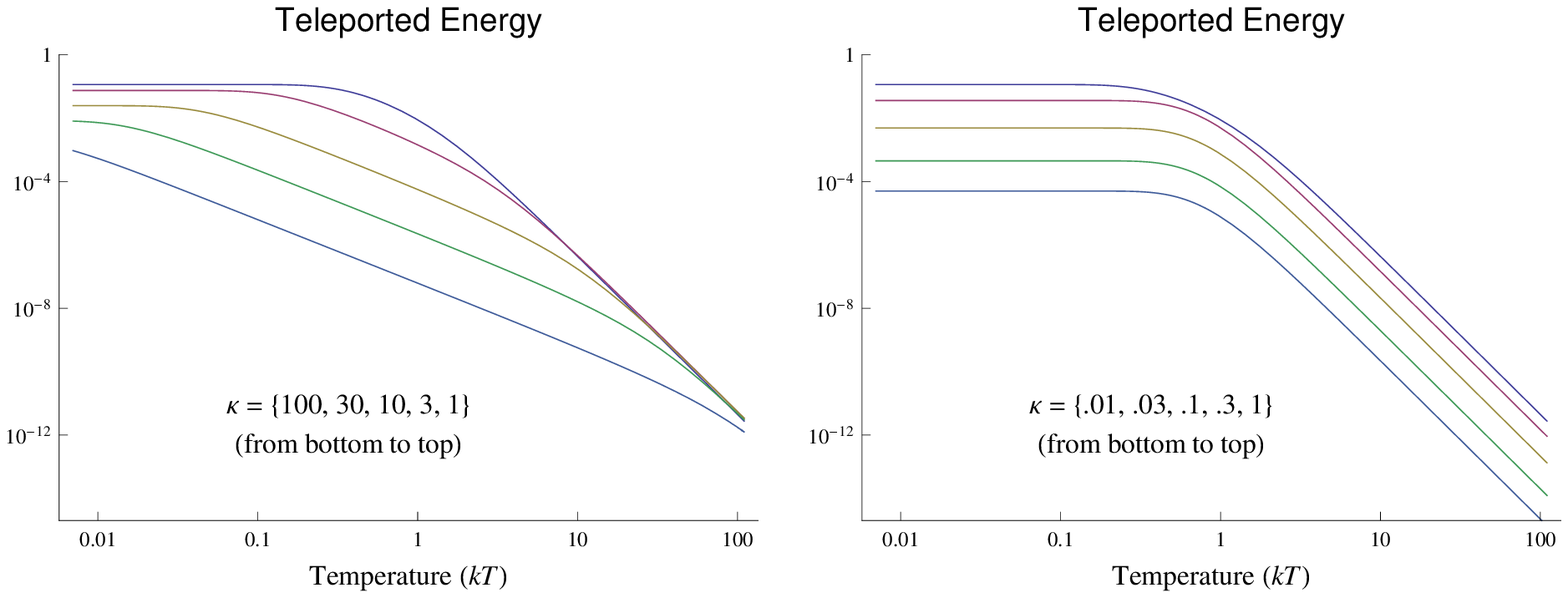}}}
\vspace*{5mm}
\begin{center}
\parbox{3.45in}{\small
Figure 2. Teleported energy $E_B(\theta_o)$ extracted at the site of particle B. In
correspondence with Fig.\ 1, the left panel shows cases of the particle pair model
with coupling $\kappa \geq 1$, and the right panel shows cases with $\kappa \leq 1$.}
\end{center}
\vspace*{0.3in}
\end{figure}

\begin{center}
{\large\bf IV. ENERGY EXTRACTION WITHOUT TELEPORTATION}
\end{center}
\vspace*{-2mm}
 
We have shown that the QET protocol extracts energy from particle B. One might ask
whether energy could as well be extracted from B directly by a local one-qubit
operation without the exercise of the QET protocol. To answer this question and
better understand what transpires in energy teleportation, suppose we execute a
frustrated version of the protocol in which no measurement of particle A is made and,
therefore, nothing is communicated to the site of B. In other words, suppose we skip
steps I and II of the protocol and just perform a conditionally unitary operation
locally on particle B, consistent with step III of the protocol. A conditionally
unitary operation made locally on B takes the form ${\cal I}\otimes{\cal W}$
where ${\cal I}$ is the identity operation (on A) and the operation ${\cal W}$
on a qubit in state $\tau$ takes the form
\begin{equation}
{\cal W}(\tau) = \sum_k p_k \mbox{\bf W}_k \tau \mbox{\bf W}_k^{\dagger}
\label{eq:uni}
\end{equation}
\noindent
where each operator $\mbox{\bf W}_k$ is unitary. We allow any number of unitary
operators $\mbox{\bf W}_k$ in (\ref{eq:uni}), with any probabilities $p_k$ that
are independent of the particles' history such that $\sum_k p_k = 1$, and we
seek the operation ${\cal W}$ that extracts the most energy possible when
applied to particle B of the particle pair in state $\rho(T)$. A general
one-qubit unitary operator is \cite{niel}
\begin{displaymath}
\mbox{\bf W} = \left( \matrix{ e^{i\frac{u}{2}}\cos\frac{w}{2}
& -e^{-i\frac{v}{2}}\sin\frac{w}{2} \cr
e^{i\frac{v}{2}}\sin\frac{w}{2} & e^{-i\frac{u}{2}}\cos\frac{w}{2} } \right)
\end{displaymath}
\noindent
with real angles $u,v,w$. Consider the operation ${\cal W}$ in (\ref{eq:uni}) with
just the single operator $\mbox{\bf W}_1=\mbox{\bf W}$; i.e., $p_1 = 1$. We find
after some calculation that, after application of ${\cal I}\otimes{\cal W}$ for
this case, the average energy in the particle pair is
\begin{eqnarray}
&& \!\!\!\!\!\!\!\!\!\!\!\!
\mbox{tr} [\mbox{\bf H}(\mbox{\bf I}\otimes\mbox{\bf W})\rho(T)(\mbox{\bf I} \otimes
\mbox{\bf W})^\dagger ] = 2m - r(1+\cos w) + \kappa c_1\cos v (1-\cos w) \nonumber \\
&& \;\;\;\;\;\;\;\;\;\;\;\;\;\;\;\;\;\;\;\;\;\;\;\;\;\;\;\;\;\;\;\;\;\;\;\;
\;\;\;\;\;\;\; -\, \kappa c_1 \cos u (1+\cos w)  \, .
\label{eq:post}
\end{eqnarray}
We seek values of the angles $u,v,w$ of $\mbox{\bf W}$ to minimize (\ref{eq:post}) and thereby
extract the maximum amount of energy with $\mbox{\bf W}$. Because $r,c_1>0$ we easily see
that (\ref{eq:post}) is minimum uniquely when $u=w=0$; that is, when
$\mbox{\bf W}=\mbox{\bf I}$, in which case (\ref{eq:post}) is exactly (\ref{eq:h}) and
zero energy is extracted. This result exmplifies Gibbs states' general property of passivity
\cite{pusz}, whereby any nontrivial unitary operation necessarily increases the
system energy and no unitary operation on $B$ can extract energy, at any temperature.
When ${\cal W}$ in (\ref{eq:uni}) is a non-trivial sum involving more than one unitary
operator, the particle pair energy is minimized when each $\mbox{\bf W}_k = \mbox{\bf I}$.
Thus, no local, conditionally unitary operation made on particle B, made without a measurement
of A and the knowledge therefrom, can do better than extract zero energy. In the QET protocol
the measurement of particle A both ``sets'' particle B and provides information to the site
of B for exploiting that setting. This is the essence of QET.

No conditional unitary operation applied locally to particle B without preparation
at and communication from A can extract energy from B. With the QET protocol, on the
other hand, energy can be extracted locally from B. This comparison, energy teleportation
versus a local conditionally unitary operation, seems most apt since step III in energy
teleportation is one of two unitary operations, the choice depending on information
sent from A to B. Conditionally unitary operations are not the most general one-qubit
operations, of course. One might ask about how quantum energy teleportation fares in
contest with a general quantum operation applied locally to B. We explore this question
now.

A quantum operation for a qubit in state $\tau$ is, in general operator-sum
form \cite{niel},
\begin{equation}
{\cal G}(\tau) = \sum_{k=1}^4 \mbox{\bf K}_k \tau \mbox{\bf K}_k^{\dagger}
\end{equation}
with Kraus operators
\begin{equation}
\mbox{\bf K}_k = \left(  \matrix{ s_k & t_k \cr u_k & v_k  }
\right)
\end{equation}
\noindent
whose complex-valued elements $s_k, t_k, u_k, v_k$ satisfy the completeness condition
\begin{equation}
\sum_{k=1}^4 \mbox{\bf K}_k^\dagger \mbox{\bf K}_k = \mbox{\bf I} \, .
\label{eq:compl}
\end{equation}
\noindent
This condition can be expressed in terms of the elements of the $\mbox{\bf K}_k$ as
\begin{eqnarray}
&& \mbox{\bf s}^\dagger \mbox{\bf s} + \mbox{\bf u}^\dagger \mbox{\bf u} = 1  \nonumber \\
&& \mbox{\bf t}^\dagger \mbox{\bf t} + \mbox{\bf v}^\dagger \mbox{\bf v} = 1  \label{eq:compl2} \\
&& \mbox{\bf s}^\dagger \mbox{\bf t} + \mbox{\bf u}^\dagger \mbox{\bf v} = 0  \nonumber
\end{eqnarray}
\noindent
where $\mbox{\bf s}=(s_1 \; s_2 \; s_3 \; s_4)^\top$, etc. We call a vector
$\mbox{\bf z}=(\mbox{\bf s}^\top,\mbox{\bf t}^\top,\mbox{\bf u}^\top,\mbox{\bf v}^\top)^\top$
feasible and write $\mbox{\bf z} \in {\cal F}$ if $\mbox{\bf z}$ satisfies the conditions in
(\ref{eq:compl2}). Using these conditions, we find after some calculation that
\begin{eqnarray}
\mbox{tr} [\mbox{\bf H}\sum_k (\mbox{\bf I}\otimes\mbox{\bf K}_k)\rho(T)(\mbox{\bf I}
\otimes\mbox{\bf K}_k)^\dagger ] &&
\!\!\!\!\!\!\!\!\!\! = 2m-2r-(1-r) \mbox{\bf u}^\dagger \mbox{\bf u}
+ (1+r)\mbox{\bf t}^\dagger \mbox{\bf t}     \nonumber \\[-2mm]
&& -\, \kappa c_1 (\mbox{\bf s}^\dagger \mbox{\bf v} + {\mathbf v}^\dagger \mbox{\bf s}
+ \mbox{\bf u}^\dagger \mbox{\bf t} + \mbox{\bf t}^\dagger \mbox{\bf u})     \; .
\label{eq:geny}
\end{eqnarray}
\noindent
Then, subtracting (\ref{eq:geny}) from (\ref{eq:h}), we find, for any $\kappa$ and
$T$, that the energy extracted by applying $\cal G$ locally to B is
\begin{equation}
\Omega(\mbox{\bf z}) = (1-r) \mbox{\bf u}^\dagger \mbox{\bf u} - (1+r)\mbox{\bf t}^\dagger \mbox{\bf t}
+\kappa c_1 (\mbox{\bf s}^\dagger \mbox{\bf v} + {\mathbf v}^\dagger \mbox{\bf s}
+ \mbox{\bf u}^\dagger \mbox{\bf t}  + \mbox{\bf t}^\dagger \mbox{\bf u}-2)   \, .
\label{eq:Om}
\end{equation}
\noindent
To find the maximum energy that can be extracted locally by ${\cal G}$, we maximize
(\ref{eq:Om}) subject to (\ref{eq:compl2}). The solution space for this problem has 32
real parameters (corresponding to the 16 complex elements of the Kraus operators), with
non-trivial constraints imposed by (\ref{eq:compl2}). Despite the evident challenge,
a simple analytical expression for the maximum can be given. The maximum of
$\Omega(\mbox{\bf z})$ subject to (\ref{eq:compl2}) is
\begin{equation}
\max\limits_{\mbox{\bf z} \in {\cal F}}  \Omega(\mbox{\bf z}) =   \left\{
\matrix{ \sqrt{\frac{1-r^2+4\kappa^2 c_1^2}{1-r^2}} -2\kappa c_1 - r \, ,
& \kappa c_1 < \frac{1-r^2}{2r} \cr   0 \, , & \mbox{otherwise} }    \right.      \; .
\label{eq:maxxx}
\end{equation}
\noindent
This maximum has two branches, a zero branch obtained by the identity operation and a
positive branch obtained by the local quantum operation ${\cal G}$ with Kraus operators
\begin{equation}
\mbox{\bf K}_1 = \left(  \matrix{ \cos\alpha & 0 \cr 0 & \cos\beta  } \right) \, , \;\;
\mbox{\bf K}_2 = \left(  \matrix{ 0 & \sin\beta  \cr \sin\alpha & 0  } \right) \, , \;\;
\mbox{\bf K}_3 = \mbox{\bf K}_4 = \left(  \matrix{ 0 & 0 \cr 0 & 0} \right)
\end{equation}
\noindent
with angles $\alpha$ and $\beta$ whose sum $\sigma=\alpha+\beta$ and difference
$\delta=\alpha-\beta$ are given by
\begin{equation}
\cos\sigma=\frac{2\kappa c_1 r}{1-r^2} \, , \;\;
\cos\delta=\frac{2\kappa c_1}{\sqrt{(1-r^2)(1-r^2+4\kappa^2 c_1^2)}}  \, . \label{eq:angles}
\end{equation}

To prove (\ref{eq:maxxx}), let $\mbox{\bf z}=
(\mbox{\bf s}^\top,\mbox{\bf t}^\top,\mbox{\bf u}^\top,\mbox{\bf v}^\top)^\top$
be any feasible vector $\mbox{\bf z} \in {\cal F}$ and associate to $\mbox{\bf z}$
the vector $\mbox{\bf z}_o= (\mbox{\bf s}_o^\top,
\mbox{\bf t}_o^\top,\mbox{\bf u}_o^\top,\mbox{\bf v}_o^\top)^\top$ with
\begin{eqnarray}
&& \mbox{\bf s}_o = (s,0,0,0)^\top  \, , \;\;\;  \mbox{\bf t}_o = (0,t,0,0)^\top  \nonumber \\
&& \mbox{\bf u}_o = (0,u,0,0)^\top  \, , \;\;  \mbox{\bf v}_o = (v,0,0,0)^\top  \nonumber
\end{eqnarray}
\noindent
where $s= \sqrt{\mbox{\bf s}^\dagger \mbox{\bf s}}$,
$t= \sqrt{\mbox{\bf t}^\dagger \mbox{\bf t}}$,
$u= \sqrt{\mbox{\bf u}^\dagger \mbox{\bf u}}$ and
$v= \sqrt{\mbox{\bf v}^\dagger \mbox{\bf v}}$ are
the magnitudes of $\mbox{\bf s}$, $\mbox{\bf t}$, $\mbox{\bf u}$ and $\mbox{\bf v}$.
This association of $\mbox{\bf z}_o$ with $\mbox{\bf z}$ is the proof's key step. We
have $\mbox{\bf z}_o \in {\cal F}$. Also,
$\mbox{\bf u}^\dagger \mbox{\bf u}=u^2=\mbox{\bf u}_o^\dagger \mbox{\bf u}_o$ and
$\mbox{\bf t}^\dagger \mbox{\bf t}=t^2=\mbox{\bf t}_o^\dagger \mbox{\bf t}_o$ and,
by the Cauchy-Schwarz inequality,
$\mbox{\bf s}^\dagger \mbox{\bf v}+\mbox{\bf v}^\dagger \mbox{\bf s}\leq 2sv
=\mbox{\bf s}_o^\dagger \mbox{\bf v}_o+\mbox{\bf v}_o^\dagger \mbox{\bf s}_o$ and
$\mbox{\bf u}^\dagger \mbox{\bf t}+\mbox{\bf t}^\dagger \mbox{\bf u}\leq 2ut
=\mbox{\bf u}_o^\dagger \mbox{\bf t}_o+\mbox{\bf t}_o^\dagger \mbox{\bf u}_o$.
Using these results with $0\leq r\leq 1$ and $c_1\geq 0$, we have
$\Omega(\mbox{\bf z}) \leq \Omega(\mbox{\bf z}_o)$. Therefore, the maximum of
$\Omega(\mbox{\bf z})$ subject to the conditions of (\ref{eq:compl2}) equals the maximum of
\begin{displaymath}
\omega(s,t,u,v) = (1-r)u^2-(1+r)t^2+2\kappa c_1(sv+ut-1)
\end{displaymath}
\noindent
subject to $s^2+u^2=1$ and $t^2+v^2=1$. Set
\begin{eqnarray}
&& s=\cos\alpha \, , \;\;\; t=\sin\beta \, , \nonumber \\
&& u=\sin\alpha \, , \;\;\; v=\cos\beta \, , \nonumber
\end{eqnarray}
and let $\sigma = \alpha+\beta$ and $\delta = \alpha - \beta$. Then
$\omega(s,t,u,v)=\varpi(\sigma,\delta)$ where
\begin{equation}
\varpi(\sigma,\delta)=\sin\sigma\cos\delta-r\cos\sigma\sin\delta-2\kappa c_1\sin\delta
\label{eq:varpi}
\end{equation}
and the maximum of $\Omega(\mbox{\bf z})$ subject to constraints (\ref{eq:compl2}) is the
unconstrained maximum of $\varpi(\sigma,\delta)$. Solving for the stationary point(s) of
$\varpi(\sigma,\delta)$ yields (\ref{eq:angles}) for $2\kappa c_1 r<1-r^2$ and
$\sigma = \delta = 0$ otherwise. Substituting (\ref{eq:angles}) into (\ref{eq:varpi})
gives the positive branch of (\ref{eq:maxxx}); putting $\sigma = \delta = 0$ into
(\ref{eq:varpi}) gives the zero branch of (\ref{eq:maxxx}).

Two cases, $\kappa = 0$ and $T=\infty$, of (\ref{eq:Om}) have particular physical interest
and can be solved directly to independently check (\ref{eq:maxxx}). In both cases
$\rho(T)$ has zero discord and the QET protocol yields zero energy.

\underline{$\kappa = 0$}: This is the case of uncoupled spin-$\frac{1}{2}$ particles in
the product state $\rho(T)=\rho_A\otimes\rho_B$ where $\rho_A=\rho_B=\frac{1}{2}
(\mbox{\bf I}-r \mbox{\boldmath$\sigma$}_z)$ with $r=\tanh\frac{1}{kT}$. We are in
this case effectively just seeking the maximum amount of energy that can be extracted
from a thermal qubit with Hamiltonian $\mbox{\bf H}_B = \mbox{\bf I}
+ \mbox{\boldmath$\sigma$}_z$ as in (\ref{eq:comp}). For this case (\ref{eq:Om}) is just
\begin{equation}
\Omega(\mbox{\bf z}) = (1-r) \mbox{\bf u}^\dagger \mbox{\bf u}
- (1+r)\mbox{\bf t}^\dagger \mbox{\bf t} \; ,
\label{eq:kap}
\end{equation}
\noindent
and, subject to (\ref{eq:compl2}), the maximum of (\ref{eq:kap}) can be seen by
inspection to be $1-r$. The energy (\ref{eq:h}) initially in the two uncoupled
particles is $2-2r$, half associated with each particle. We conclude that in the
case $\kappa = 0$ the optimal local quantum operation ${\cal G}$ extracts all
the energy $1-r$ associated with particle B. This agrees with (\ref{eq:maxxx});
for $\kappa=0$ we have $q=0$ and the positive branch of (\ref{eq:maxxx}) applies,
yielding $1-r$ for the energy locally available at B. There is no non-local
energy in A and B in this example.

\underline{$T =\infty$}: Again the particles are uncoupled, with in this case
$\rho(\infty) = \frac{1}{4}\mbox{\bf I}\otimes \mbox{\bf I}$ with energy
$\mbox{tr} [\mbox{\bf H} \rho(\infty)] = 2m$. Also, $r=c_1=0$ so
\begin{equation}
\Omega(\mbox{\bf z})= \mbox{\bf u}^\dagger \mbox{\bf u}-\mbox{\bf t}^\dagger \mbox{\bf t}  \; .
\label{eq:inft}
\end{equation}
\noindent
The maximum of (\ref{eq:inft}) subject to (\ref{eq:compl2}) is $1$ by inspection.
This maximum energy extractable locally by $\cal G$ agrees with (\ref{eq:maxxx});
$r=q=0$ for $T=\infty$ so the positive branch of (\ref{eq:maxxx}) applies with
maximum value 1. The non-local energy in A and B is $2m-2$ in this example.

\begin{figure}[b!]
\vspace*{7mm}
\centerline{\scalebox{1}{\includegraphics{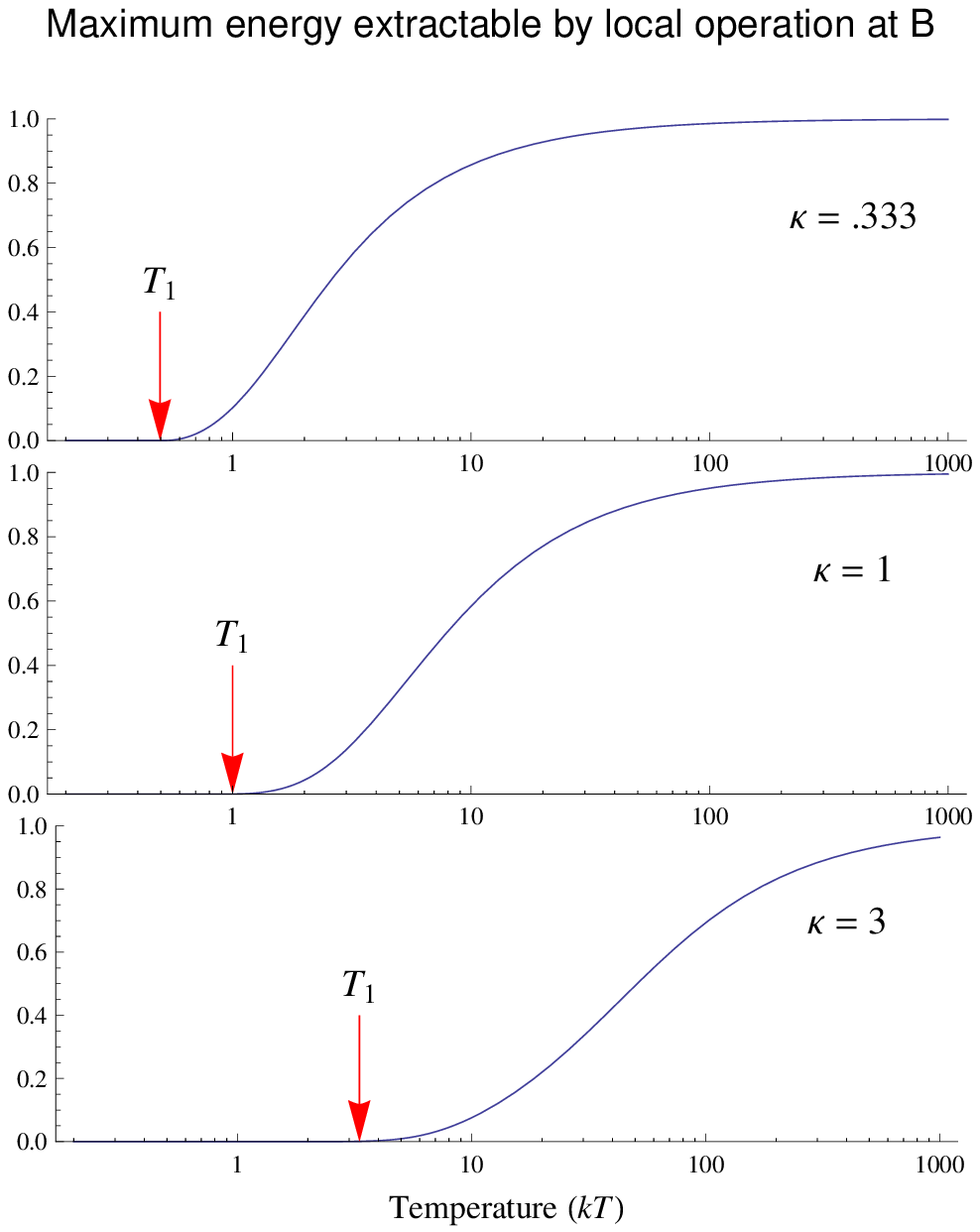}}}
\vspace*{7mm}
\begin{center}
\parbox{4.0in}{\small
Figure 3. The maximum energy that can be extracted at the site of particle B
by a local quantum operation ${\cal G}$. Below the temperature threshold $T_1$
no energy can be extracted by any ${\cal G}$.}
\end{center}
\vspace*{-0.1in}
\end{figure}

In terms of our state parameters $\kappa$ and $T$, the condition
$2\kappa r c_1 < 1-r^2$ in (\ref{eq:maxxx}) is equivalent to
\begin{equation}
\left( \kappa \sinh\frac{2m}{kT} +m \sinh\frac{2\kappa}{kT} \right)^2
< 2m^2\cosh\frac{2\kappa}{kT} \left( \cosh\frac{2m}{kT}+ \cosh\frac{2\kappa}{kT}\right)  \, .
\label{eq:first}
\end{equation}
\noindent
A unique non-zero temperature threshold $T_1$ saturates condition (\ref{eq:first}) for each
coupling $\kappa$. Below temperature $T_1$ no energy can be extracted at site B by any local
operation $\cal G$. Figure 3 shows this threshold for different couplings $\kappa$. The
emergence of this threshold in our simple two-particle model is unexpected. Its existence
points to a distinctive ability of energy teleportation. Below $T_1$ energy can be extracted
at site B {\em only} by QET; {\em no} local operation, unitary or otherwise, at site B can
accomplish this.

\begin{figure}[t]
\centerline{\scalebox{1}{\includegraphics{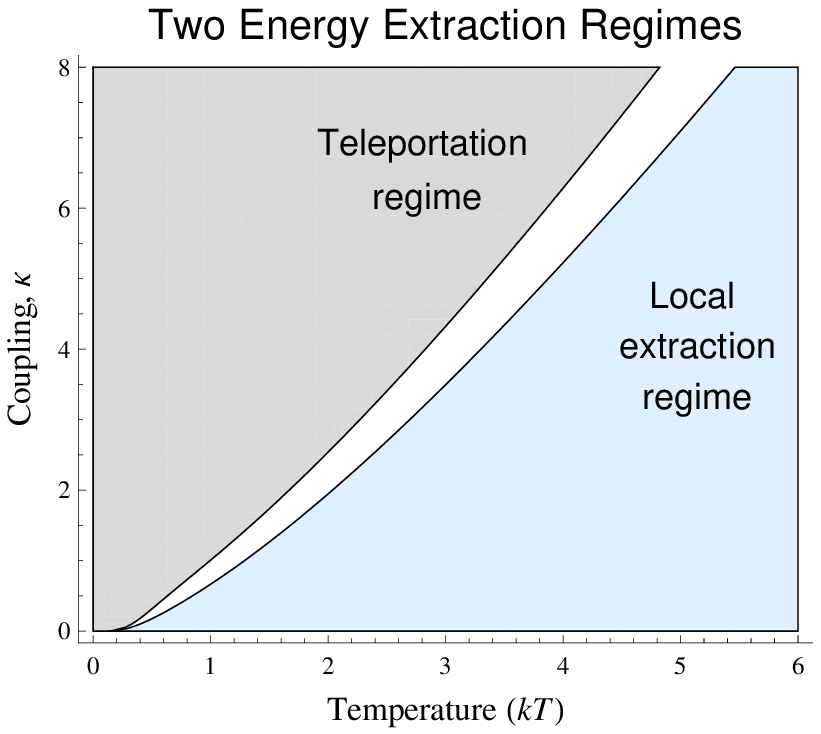}}}
\vspace*{5mm}
\begin{center}
\parbox{5in}{\small
Figure 4. Two temperature-dependent energy regimes. QET yields energy throughout the
$\kappa$--$T$ parameter space; in the teleportation regime energy extraction is
possible {\em only} by QET. In the local extraction regime a local operation yields more
energy than QET. In the narrow temperature window $(T_1,T_2)$ between the two regimes,
QET yields more energy than available by a local operation.}
\end{center}
\vspace*{.1in}
\end{figure}

\begin{figure}[t]
\centerline{\scalebox{1}{\includegraphics{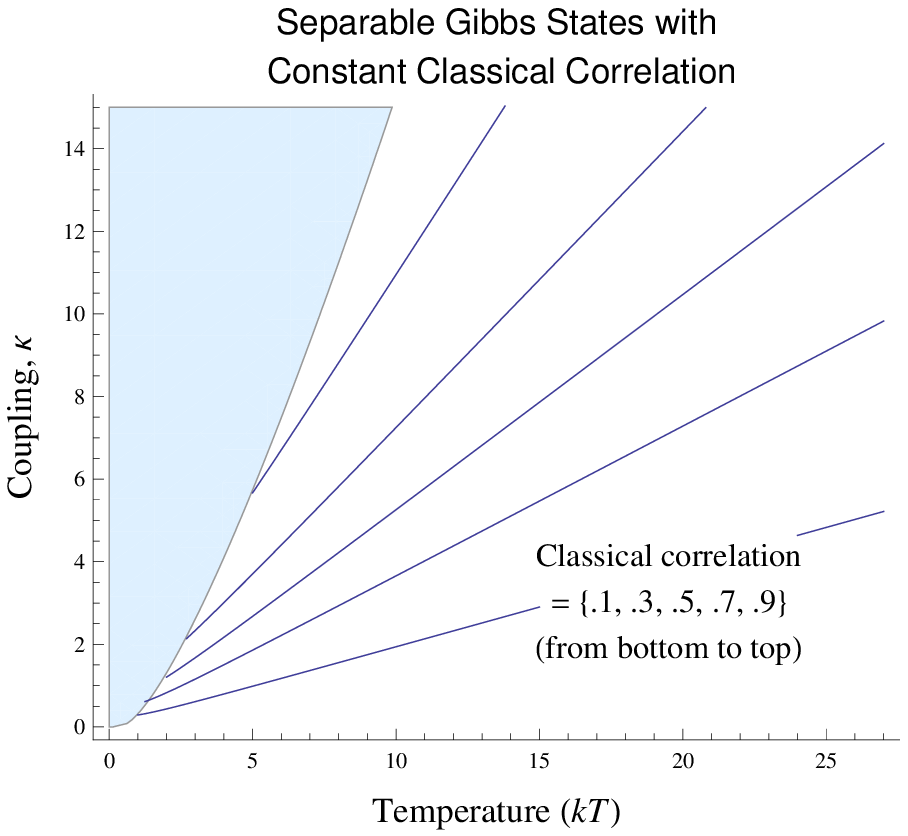}}}
\vspace*{10mm}
\begin{center}
\parbox{5in}{\small
Figure 5. $\kappa$--$T$ parameter space of two-particle Gibbs states. Gibbs states in the
shaded region are entangled. The contours in the unshaded region are parametric families $\rho_c(T)$ of separable
Gibbs states with constant classical correlation, $C[\rho_c(T)] = .1,\, .3,\, .5,\, .7,\, .9$.}
\end{center}
\vspace*{0.1in}
\end{figure}

Above $T_1$ is a second temperature threshold $T_2$, defined by
$\max_{\mbox{\bf z}}\Omega(\mbox{\bf z})=E_B$. This is the temperature where the energies
available at particle B by teleportation and by local operation are equal. In the
temperature window $(T_1,T_2)$ a local operation on B can extract energy, but not as
much as that yielded by energy teleportation. Above temperature $T_2$, a local operation
can yield more energy than teleportion. Returning to Fig.\ 2, we can check that for any
coupling $\kappa$ the amount of energy teleported at temperatures above $T_2$ is vanishingly
small relative to the amount available by $\cal G$. We conclude therefrom that there are
effectively two energy regimes: a teleportation regime below the window $(T_1,T_2)$ where
energy can be extracted from B by teleportation but not by any local operation $\cal G$ and a
local extraction regime where energy can be extracted by $\cal G$ but very little energy
can be teleported. These two energy regimes are shown in Fig.\ 4.

\begin{figure}[t]
\centerline{\scalebox{1}{\includegraphics{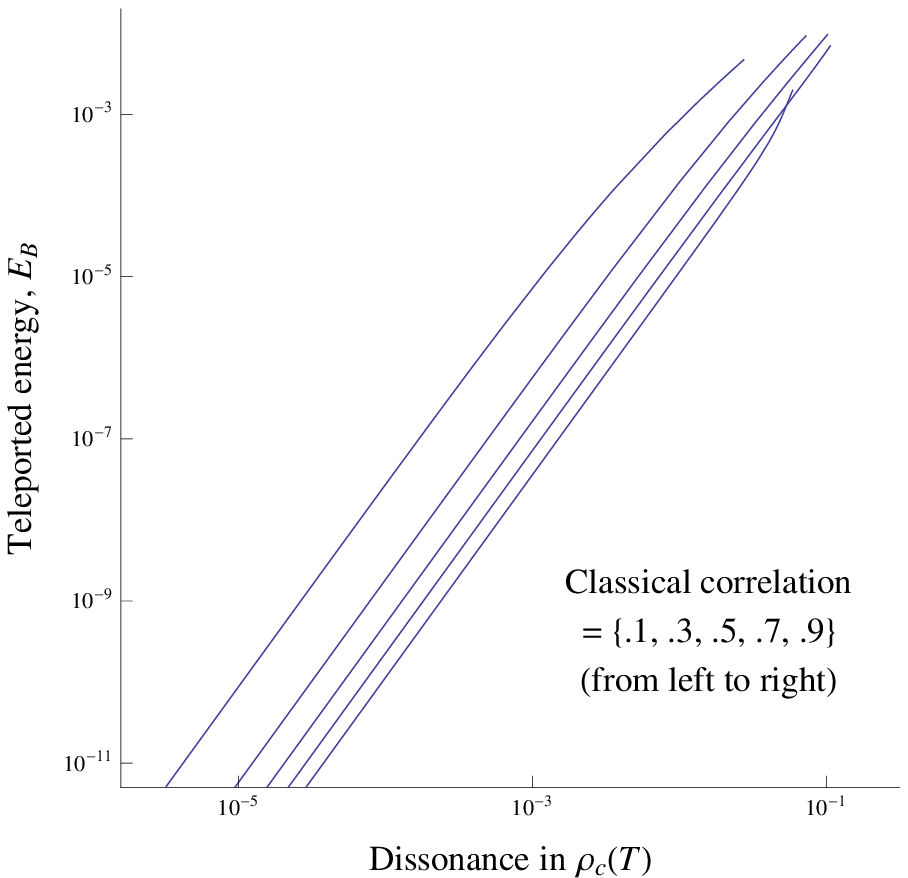}}}
\vspace*{8mm}
\begin{center}
\parbox{5in}{\small
Figure 6. Parametric plot of teleported energy $E_B$ versus thermal discord $D[\rho_c(T)]$
for varying temperature for five levels of classical correlation. Each curve
shows the correspondence between quantum dissonance and teleported energy at
temperatures above $T_e$ where $\rho_c(T)$ is separable.}
\end{center}
\vspace*{0.2in}
\end{figure}

\vspace*{-1mm}
\begin{center}
{\large\bf V. DISCUSSION AND SUMMARY}
\end{center}
\vspace*{-4mm}

Product states cannot support QET. The two cases, $\kappa=0$ and $T=\infty$, of the
previous section support this---in each case $\rho(T)$ is a product state and the
teleported energy is zero---and one can readily see that this is generally true. In our
two-particle model QET requires some correlation between the particles at the two sites.
By definition, the quantum part of this correlation is measured by the quantum discord
in the particles' state. We found for all finite temperatures and non-zero degrees of
spin coupling both that quantum correlation is present in $\rho(T)$ and that, in fact,
energy teleportation can be accomplished with the QET protocol. The quantum correlation
is solely entanglement at zero temperature. At temperatures above $T_e$ (marked by the
points on the curves in Fig.\ 1), $\rho(T)$ is separable, the quantum correlation that
exists between particles A and B is strictly dissonant, and this dissonance is acting
as a quantum resource for energy teleportation. This means not just that energy
teleportation occurs when dissonance is present in $\rho(T)$. It means that quantum
states with more dissonance yield greater teleported energy, and that this occurs under
controlled conditions in which all other correlations are held constant (and cannot,
therefore, account for the greater energy.) At temperatures above $T_e$ where $\rho(T)$
is strictly dissonant, $\rho(T)$ has for fixed $\kappa$ a varying amount of classical
correlation $C[\rho(T)]$ that could conceivably be the source of energy increases. To
eliminate $C[\rho(T)]$ as a possible factor, we consider $\rho(T)$ as a two-parameter
($T$ and $\kappa$) family of states and choose (as in Fig.\ 5) parametric families $\rho_c(T)$
of Gibbs states with constant $C[\rho_c(T)]$---accomplished by adjusting $\kappa$ to keep
$C[\rho(T)]$ constant as $T$ is increased. When this is done, we get the results in Fig.\ 6.
The curves in Fig.\ 6 are parametric plots of teleported energy versus the quantum
dissonance in $\rho_c(T)$ as temperature is increased starting at $T_e$ and the coupling
$\kappa$ is varied to maintain constant $C[\rho(T)]$. The curves correspond to five fixed
values of $C[\rho_c(T)]$, chosen to represent the 0--1 range of classical correlation
present in the separable Gibbs states in Fig.\ 5. On each curve in Fig.\ 6, both entanglement
and classical correlation are fixed (to zero in the case of entanglement); only the
dissonance is varying. These curves show, with all other forms of correlation held
constant, a direct relationship between $E_B$ and the Gibbs state dissonance: greater
dissonance consistently yields greater energy. Dissonance is acting in the case of the Gibbs
states studied here as a quantum resource. Other examples of dissonance as a resource are
known \cite{roa,datt,mod2,frey}; more examples, we expect, will contribute to a fuller
understanding of dissonance, discord and quantum correlation.

When we consider a strictly local quantum operation to extract energy from particle B of
our spin pair, we find the temperature threshold $T_1$ below which no energy extraction
is possible. That such a threshold emerges in even our very simple two-particle model is
remarkable. We refer to temperatures below $T_1$ as the teleportation regime because at
these temperatures the QET protocol unlocks energy that is otherwise not locally accessible.
We also identified a threshold $T_2 > T_1$ above which more energy can be extracted by a
local operation than by energy teleportation. In fact, in the spin system we consider, energy
teleportation yields only an vanishingly small amount of energy above $T_2$, while at these
temperatures significant energy can be extracted by local operation on B. We therefore call
these temperatures above $T_2$ the local extraction regime. Between these two identified
regimes is a narrow transitional temperature window. It would interesting to know whether,
and to what extent, these regimes exist in other spin chains. In particular, we would like
to know whether a teleportation regime with an associated $T_1$ threshold similar to that
identified here is generic to different classes of spin chains.

\vspace*{6mm}
\begin{center}
{\large\bf ACKNOWLEDGMENT}
\end{center}

M.H. would like to thank A. Shimizu, T. Sagawa, K. Funo, H. Tasaki and
Y. Watanabe for illuminating conversations about quantum thermal states.
\vspace*{3mm}

\newpage
\begin{center}
{\large\bf APPENDIX A}
\end{center}
\vspace*{-2mm}

The classical correlation (\ref{eq:clas}) in the Gibbs state $\rho(T)$ is
\begin{equation}
C[\rho(T)] = h(r) - \min\limits_{\{\mbox{\scriptsize\bf M}_k\}}
\sum_k q_k S(\rho_k)
\label{eq:clascorr}
\end{equation}
\noindent
where, with $h(x)$ as in (\ref{eq:hx}), $h(r)=S(\rho_A(T))$ is the von Neumann entropy of
$\rho_A(T)=\mbox{tr}_B[\rho(T)]$. The projective measurement
$\{\mbox{\bf M}_0,\mbox{\bf M}_1\}$ of qubit B has the
general form $\mbox{\bf M}_k=|k^\prime\rangle\langle k^\prime |$ where
\begin{displaymath}
|0^\prime\rangle = \cos\frac{\theta}{2}|0\rangle +e^{i\phi} \sin\frac{\theta}{2}|1\rangle \, ,
\end{displaymath}
\begin{displaymath}
|1^\prime\rangle = \sin\frac{\theta}{2}|0\rangle -e^{i\phi} \cos\frac{\theta}{2}|1\rangle \, .
\end{displaymath}
\noindent
The measurement $\{\mbox{\bf M}_0,\mbox{\bf M}_1\}$ changes $\rho(T)$ to one
of two states
\begin{displaymath}
\rho_k = \frac{1}{q_k}(\mbox{\bf I}\otimes\mbox{\bf M}_k)\rho(T)
(\mbox{\bf I}\otimes\mbox{\bf M}_k)=  \frac{1}{q_k}\mbox{\bf J}_k\otimes\mbox{\bf M}_k \, ,
\;\;\; k=0,1
\end{displaymath}
\noindent
where
\begin{displaymath}
\mbox{\bf J}_0 = \frac{q_0}{2} \mbox{\bf I}
-\frac{c_1\cos\phi\sin\theta}{4} \mbox{\boldmath$\sigma$}_x
+\frac{c_2\sin\phi\sin\theta}{4} \mbox{\boldmath$\sigma$}_y
-\frac{r-c_3\cos\theta}{4} \mbox{\boldmath$\sigma$}_z   \, ,
\end{displaymath}
\begin{displaymath}
\mbox{\bf J}_1 = \frac{q_1}{2} \mbox{\bf I}
+\frac{c_1\cos\phi\sin\theta}{4} \mbox{\boldmath$\sigma$}_x
-\frac{c_2\sin\phi\sin\theta}{4} \mbox{\boldmath$\sigma$}_y
-\frac{r+c_3\cos\theta}{4} \mbox{\boldmath$\sigma$}_z
\end{displaymath}
\noindent
with corresponding probabilities
\begin{displaymath}
q_0(\theta) = \mbox{tr}[(\mbox{\bf I}\otimes\mbox{\bf M}_0)\rho(T)
(\mbox{\bf I}\otimes\mbox{bf M}_0)] = \frac{1-r\cos\theta}{2} \, ,
\end{displaymath}
\begin{displaymath}
q_1(\theta) = \mbox{tr}[(\mbox{\bf I}\otimes\mbox{\bf M}_1)\rho(T)
(\mbox{\bf I}\otimes\mbox{bf M}_1)] = \frac{1+r\cos\theta}{2} \, .
\end{displaymath}
\noindent
The eigenvalues of $\rho_0$ and $\rho_1$ are, respectively,
\begin{equation}
0, 0, \frac{1}{2}\pm \frac{\sqrt{A(\theta ,\phi )-2rc_3\cos\theta}}{4q_0}
\label{eq:eig11}
\end{equation}
\noindent
and
\begin{equation}
0, 0, \frac{1}{2}\pm \frac{\sqrt{A(\theta ,\phi )+2rc_3\cos\theta}}{4q_1}
\label{eq:eig22}
\end{equation}
\noindent
where
\begin{displaymath}
A(\theta ,\phi ) = \left( c_1^2\cos^2\phi +c_2^2\sin^2\phi \right)
\sin^2\theta+c_3^2\cos^2\theta + r^2 \, .
\end{displaymath}

The minimum in (\ref{eq:clascorr}) is
\begin{equation}
\min\limits_{\{\mbox{\scriptsize\bf M}_k\}} \sum_k q_k S(\rho_k)
= \min\limits_{\theta,\phi} \left( q_0(\theta) S(\rho_o)+q_1(\theta) S(\rho_1) \right)
\label{eq:miny}
\end{equation}
\noindent
where
\begin{displaymath}
S(\rho_0) = h\left( \frac{\sqrt{A(\theta ,\phi )-2rc_3\cos\theta}}{2q_0(\theta)} \right) \, ,
\end{displaymath}
\begin{displaymath}
S(\rho_1) = h\left( \frac{\sqrt{A(\theta ,\phi )+2rc_3\cos\theta}}{2q_1(\theta)} \right)
\end{displaymath}
\noindent
because the zero eigenvalues in (\ref{eq:eig11}) and (\ref{eq:eig22}) do not
contribute to the entropies of $\rho_0$, $\rho_1$. The function $h(x)$ is decreasing
for $0< x<1$ and $c_1 > c_2$ so the minimum in (\ref{eq:miny}) is found at $\phi = 0$
for all angles $\theta$. We have then that the minimum in (\ref{eq:clascorr}) is
\begin{eqnarray}
&& \min\limits_{\{\mbox{\scriptsize\bf M}_k\}} \sum_k q_k S(\rho_k)
=  \min\limits_{\theta} \left( q_0(\theta)
h\left( \frac{\sqrt{c_1^2\sin^2\theta+(r-c_3\cos\theta)^2}}{2q_0(\theta)} \right) \right.
\nonumber \\
&& \;\;\;\;\;\;\;\;\;\;\;\;\;\;\; \;\;\;\;\;\;\;\;\;\;\;\;\;\;\;
\;\;\;\;\;\;\;\;\;\;\;\;\; + \left.
q_1(\theta) h\left( \frac{\sqrt{c_1^2\sin^2\theta+(r+c_3\cos\theta)^2}}{2q_1(\theta)} \right)
\right) \, .  \label{eq:minand}
\end{eqnarray}
The minand in (\ref{eq:minand}) has, for arbitrary values of $c_1$, $c_3$, and $r$, at
least two and sometimes more local extrema \cite{huang}. However, for the class of Gibbs
states $\rho(T)$ in (\ref{eq:ther}) with $c_1$, $c_3$, and $r$ restricted to values
given by (\ref{eq:c1}), (\ref{eq:c3}), and (\ref{eq:rr}), one can check
numerically that the minimum in (\ref{eq:minand})
is achieved by $\theta = \frac{\pi}{2}$ for all $\kappa$ and $T$. Since
$q_0(\frac{\pi}{2}) = q_1(\frac{\pi}{2}) = \frac{1}{2}$,
the minimum in (\ref{eq:minand}) is $h(\sqrt{c_1^2+r^2})$. Substituting this in
(\ref{eq:clascorr}) gives our result (\ref{eq:our}) for $C[\rho(T)]$.

\begin{center}
{\large\bf APPENDIX B}
\end{center}
\vspace*{-2mm}

The four eigenvalues of the partial transpose of $\rho(T)$ are
\begin{eqnarray}
&& \!\!\!\!\!\! \lambda_{1\pm} = \frac{e^{-\frac{2m}{kT}}}{mZ}
\left( m\cosh\frac{2\kappa}{kT} \pm \kappa\sinh\frac{2m}{kT} \right) \, , \nonumber \\
&& \!\!\!\!\!\! \lambda_{2\pm} = \frac{e^{-\frac{2m}{kT}}}{mZ}
\left( m\cosh\frac{2m}{kT} \pm \sqrt{m^2\sinh^2\frac{2\kappa}{kT}
+\sinh^2\frac{2m}{kT}} \right) \, . \nonumber
\end{eqnarray}
Eigenvalues $\lambda_{1+}, \lambda_{2+}$ are always positive. Also, $\lambda_{2-}$
is always positive since
\begin{eqnarray}
\lambda_{2-}\lambda_{2+}
&& \!\!\!\!\!\!\!\!\! = \frac{e^{-\frac{4m}{kT}}}{m^2Z^2} \left(
m^2\cosh^2\frac{2m}{kT} - m^2\sinh^2\frac{2\kappa}{kT} -\sinh^2\frac{2m}{kT}
\right)                                                              \nonumber \\
&& \!\!\!\!\!\!\!\!\! = \frac{e^{-\frac{4m}{kT}}}{m^2Z^2} \left(
\kappa^2\cosh^2\frac{2m}{kT} - m^2\sinh^2\frac{2\kappa}{kT} +1
\right)                                                              \nonumber \\
&& \!\!\!\!\!\!\!\!\! \geq \frac{4\kappa^2e^{-\frac{4m}{kT}}}{k^2 T^2 Z^2} \left(
\left(\frac{kT}{2m}\right)^2\cosh^2\frac{2m}{kT}
- \left(\frac{kT}{2\kappa}\right)^2\sinh^2\frac{2\kappa}{kT}
\right)                                                              \nonumber \\
&& \!\!\!\!\!\!\!\!\! \geq \frac{4\kappa^2e^{-\frac{4m}{kT}}}{k^2 T^2 Z^2} \left(
\left(\frac{kT}{2m}\right)^2\cosh^2\frac{2m}{kT}
- \left(\frac{kT}{2m}\right)^2\sinh^2\frac{2m}{kT}
\right)                                                          \label{eq:in5}  \\
&& \!\!\!\!\!\!\!\!\! = \frac{\kappa^2e^{-\frac{4m}{kT}}}{m^2 Z^2}
> 0    \, ,                                                            \nonumber
\end{eqnarray}

\noindent
where inequality (\ref{eq:in5}) holds because $m>\kappa$ and $\sinh x/x$ is strictly
increasing for $x>0$. Therefore, according to the PPT criterion, the thermal state
$\rho(T)$ is separable if and only if $\lambda_{1-}\geq 0$. This is the source of
condition (\ref{eq:cond}).

\setlength{\baselineskip}{5.2mm}

\end{document}